\documentclass[a4paper]{mye}
\usepackage{natbib}
\usepackage[latin1]{inputenc}
\usepackage{amsbsy}
\usepackage{amsmath}
\renewcommand{\vec}[1]{\boldsymbol{#1}}
\usepackage{graphicx}

\newcommand{\myfig}[1]{Fig.~\ref{#1}}
\newcommand{\myFig}[1]{Fig.~\ref{#1}}
\newcommand{\myeq}[1]{Eq.~(\ref{#1})}

\newcommand{\eiscat}{\mbox{EISCAT}}
\newcommand{\eg}{e.g.\ }
\renewcommand{\title}[1]{
  \begin{center}
{\LARGE#1}\vskip 1.5em    
  \end{center}}
\renewcommand{\author}[2][Q]{\textbf{#2}\protect$^{#1}\protect$}
\renewcommand{\affil}[2][]{$^{#1}$#2\par}

\begin{document}
\thispagestyle{empty}
{\title{Auroral field-aligned currents by incoherent scatter
plasma line observations in the E region}
\vskip 1em
\flushleft
\begin{center}
\author[1,2]{Ingemar~Häggström},
\author[2]{Mikael~Hedin},
\author[1]{Takehiko~Aso},
\author[2]{Asta~Pellinen-Wannberg}
\author[2]{and Assar~Westman}
\vskip 1em  
\affil[1]{National Institute of Polar Research,
    1-9-10 Kaga, Itabashi-ku,\\ Tokyo 173-8515, Japan.}
\affil[2]{Swedish Institute of Space Physics, Box 812,
    S-981 28 Kiruna, Sweden.}
\end{center}
\vskip 3em
}
  
\date{} 


\begin{abstract}
\noindent
  The aim of the Swedish-Japanese \eiscat{} campaign in
  February 1999 was to measure the ionospheric parameters inside and
  outside the auroral arcs. The ion line radar experiment was
  optimised to probe the E-region and lower F-region with as high
  a speed as possible. Two extra channels were used for the plasma line
  measurements covering the same altitudes, giving a total of 3
  upshifted and 3 downshifted frequency bands of 25~kHz each. For most
  of the time the shifted channels were tuned to 3 (both), 4 (up), 5.5
  (down) and 6.5 (both) MHz.

  Weak plasma line signals are seen whenever the radar is probing the diffuse
  aurora, corresponding to the relatively low plasma frequencies. At
  times when auroral arcs pass the radar beam, significant increases
  in return power are observed. Many cases with simultaneously up and
  down shifted plasma lines are recorded. In spite of the rather
  active environment, the highly optimised measurements
  enable investigation of the properties of the plasma
  lines.

  A modified theoretical incoherent scatter spectrum is used to
  explain the measurements. The general trend is an upgoing
  field-aligned suprathermal current in the diffuse aurora,
  There are also
  cases with strong suprathermal currents indicated by large
  differences in signal strength between up- and downshifted plasma
  lines. A full fit of the combined ion and plasma line spektra
  resulted in suprathermal electron distributions consistent with models.
\end{abstract}

\section{Introduction}
\label{sec:introduction}

The incoherent scatter spectrum consists mainly of two lines, the
widely used and relatively strong ion line and the very weak easily
forgotten broadband electron line. There is also another line present,
the plasma line, due to scattering from high frequency electron
waves, namely Langmuir waves. From the downgoing and upgoing Langmuir waves,
two plasma lines can be detected by the radar. The frequency shift
from the transmitted signal is the frequency of the scattered
Langmuir wave plus the Doppler shift caused by electron drift. Plasma lines
can be used to measure the electron drift and hence the line-of-sight
electric current. The problem in ion line analysis with the
uncertainty of the radar 
constant can be solved by the plasma line frequency
determination and when that is done 
the speed
of measurement can be significantly increased by including the plasma
line in the ion line analysis. However, since the frequency of the
plasma lines is not known beforehand, and the frequency is varying
with height, it is difficult to measure them with enough resolution.

There have been a number of reports on plasma line measurements and
their interpretation. Most of them have discussed the frequency
shift from the transmitted pulse and the scattering has mainly been
from the F-region peak, \eg \citet{showen1979rs}, \citet{kofman1993jgr} and
\citet{nilsson1996jatp}. The latter two showed also that the simple
formula for the Langmuir wave frequency,
\begin{equation}
  \label{eq:1}
  f^2=f_p^2 (1+3k^2\lambda_D^2)+f_c^2\sin^2\alpha  
\end{equation}
where $f_p$ is the plasma frequency, $k$ the wave number $f_c$ the
electron gyro frequency, $\lambda_D$ the electron Debye length and
$\alpha$ the angle between the scattering wave and the magnetic field,
is valid to within a few kHz and thus enough to set the radar system
constant. To be able to
deduce any electron drifts, or current, out of the positions of the lines 
these authors also show that \myeq{eq:1} is not sufficient, and it is
necessary to carry out more accurate
calculations. \citet{hagfors1981jgr} had also to go to further
expansions in deriving the ambient electron temperature from the plasma lines.
The fact that so many reports deal with the F region peak is due to
the altitude profile shape of the plasma line frequency, which  according to
\myeq{eq:1} will also show a peak around that height. The measurements
can thus be
made relatively easily using rather coarse height resolution but good
frequency resolution and only detect the peak frequency. Measurements
using the same strategy, but at other heights, have been made with a
chirped radar by matching the plasma line frequency height gradient and the
transmitter frequency gradient
\citep{birkmayer1986jatp,isham1993jgr}. This technique allows
determination of the Langmuir frequency with very high frequency
resolution, but do not use the radar optimally, since the chirped
pulse cannot be used for anything else than plasma line measurements.

The enhancement of
the plasma lines, which occurs in the presence of suprathermal
electron fluxes \citep{perkins1965pr}, either photoelectrons or
secondaries from auroral
electrons, has been investigated by \citet{nilsson1996anngeo}, where
they also calculate predictions of plasma line strength for different
incoherent scatter radars and altitudes. They also show that the power
of the plasma line is rather structured with respect to ambient
electron density, depending on fine structure in the suprathermal
distributions due to excitations of different atmospheric
constituents.

Incoherent scatter plasma lines in aurora are more difficult
to measure since the variations in the plasma parameters are strong
with large time and spatial gradients. Reports of auroral plasma lines in the
aurora are also more rare, and most of them are based on too coarse
time resolution
\citep{wickwar1978jgr,kofman1980jgr,oran1981jgr,valladares1988jgr},
with resolutions ranging from 30 seconds up to 20 minutes. The
enhancements over the thermal level were high, but consistent with
what could be expected of model calculations of suprathermal electron
flux. They also tried to calculate currents and electron temperature
from the frequency shifts of the plasma lines but with very large
error bars.
\citet{kirkwood1995jgr} used the \eiscat{} radar
and the filter bank technique
and recorded much higher intensities of the plasma lines, since they
got down to resolutions of 10 seconds, and showed also
that the plasma-turbulence model proposed by
\citet{mishin1994jgr} was not consistent with the data, but could be
explained by reasonable fluxes of suprathermal electrons.

In this paper we present data obtained with the high resolution
alternating code technique \citep{lehtinen1987rs}, as
was also done for F-region plasma lines by \citet{guio1996anngeo}, with even
higher intensities due to the time resolutions of 5 seconds.
An interesting, but at the time of the experiment not realisable at \eiscat,
technique would have been the type of coded long pulses used by
\citet{sulzer1994jgr} for HF-induced plasma lines.
From relative strengths between up- and downshifted lines we detect a
general trend of upgoing field-aligned currents in the diffuse aurora
carried by the suprathermal electrons. We propose a generalisation of the
theoretical incoherent scatter spectrum, to include multiple shifted electron
distributions, and in one example we do a full 7-parameter fit of the
incoherent scatter spectrum, including the enhanced plasma lines
assuming Maxwellian secondary electrons, resulting in the first radar
measurement of its flux and a current carried by the thermal
electrons.

\section{Experiment}
\label{sec:experiment}

The measurements we present were collected by the 930~MHz \eiscat{} UHF
incoherent scatter radar, with its transmitter located at Ramfjordmoen
in Norway (69.6~°N, 19.2~°E, L=6.2). The signals scattered from the
ionosphere were received at stations in Kiruna, Sweden and Sodankylä,
Finland as well as at the transmitting site. General descriptions of the
radar facility are given by \citet{folkestad1983rs} and \citet{baron1984jatp}.
Local magnetic midnight at Ramfjordmoen is at about 2130~UT.
The aim, in the Swedish-Japanese \eiscat{} campaign in February 1999,
was to measure the ionospheric parameters inside and outside
the auroral arcs.

\begin{figure}[htbp]
  \begin{center}
    \includegraphics[width=\columnwidth]{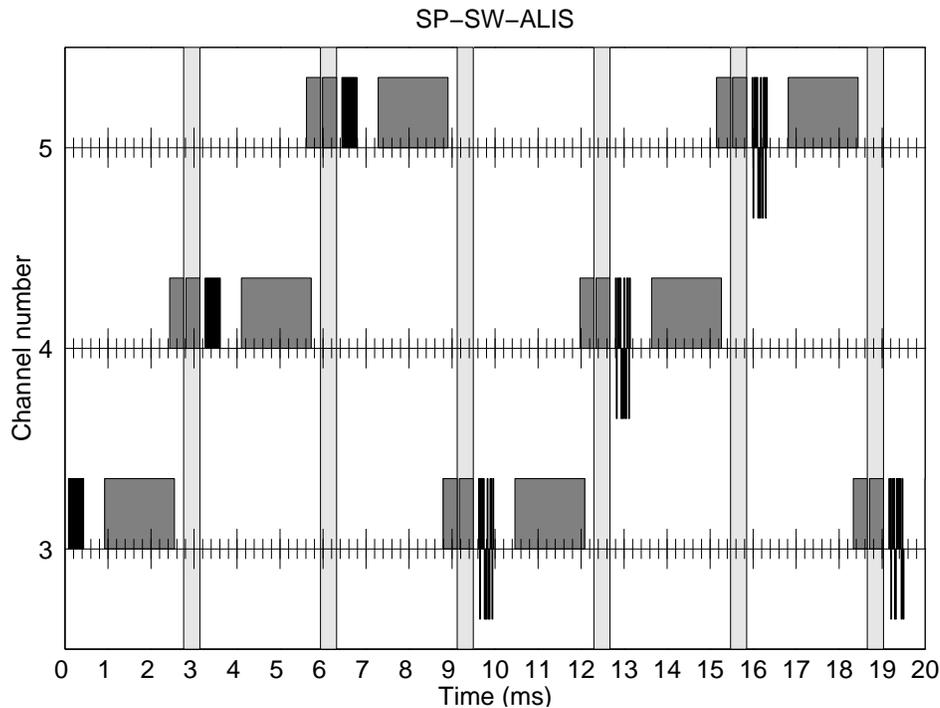}
    \caption{Transmission (black) and reception (dark gray) scheme for part of
      SP-SW-ALIS. A 16-bit alternating code, 22~$\mu$s bits, is cycled
      over 3 frequencies. The interscan period is 9~ms and the total
      cycle takes 300~ms. The plasma line channels were set to sample
      the same range extent as the ion lines.
    \label{fig:txrx}}
  \end{center}
\end{figure}

\begin{figure}[htbp]
  \begin{center}
    \includegraphics[width=\columnwidth]{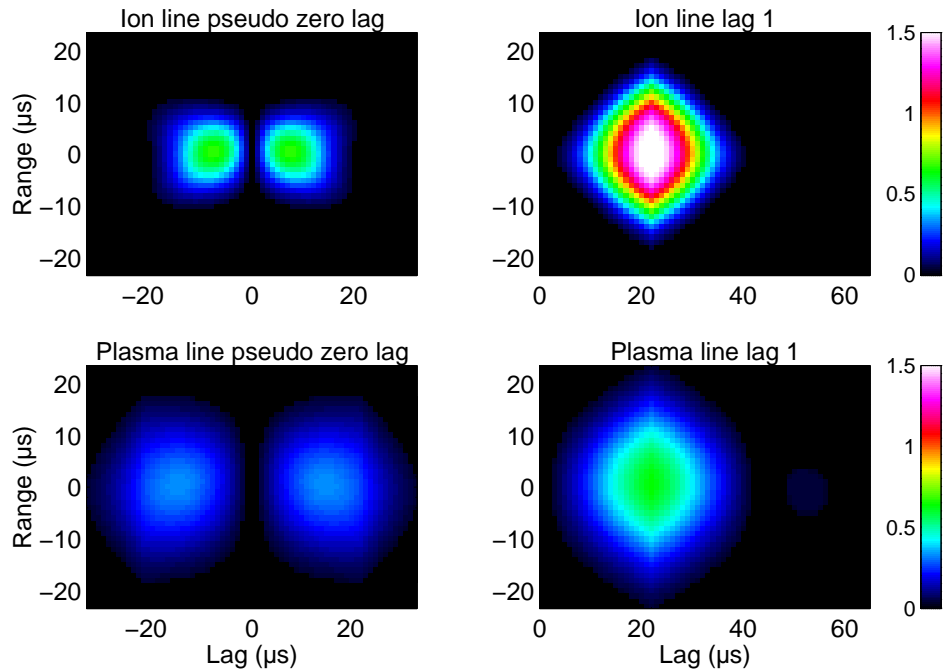}
    \caption{Range-lag ambiguity function for the first
      two lags, a) ion line and b) plasma line. The range, given in
      $\mu$s, can be converted to km with multiplication of 0.15. The
      differences between the ion and plasma line ambiguity functions
      are due to the use of different receiver filters.
    \label{fig:amb}}
  \end{center}
\end{figure}

For this a 3 channel ion line alternating code \citep{lehtinen1987rs}
experiment, optimised to probe the
E-region and lower F-region with as high a speed as possible, was
developed. The 16 bit strong condition alternating code with bitlengths of
22~$\mu$s was used, giving 3 km range resolution, and with a sample rate of
11~$\mu$s the range separations in consecutive spectra were 1.65~km.
\myFig{fig:txrx} shows the transmission/reception scheme of the first
20~ms of the radar cycle.
The whole alternating code sequence takes
about 0.3~s to complete. During this period the incoherent scatter
autocorrelation functions (ACF) at the probed heights should not
change significantly for the alternating codes to work. In order to
keep this as short as possible, the short pulses, normally used for
zerolag estimation, were dropped and instead a pseudo zero lag,
obtained from decoding the power profiles of the different codes in
the alternating code sequence, was used. \myFig{fig:amb} shows the
range-lag ambiguity function for this lag centred around 0~$\mu$s in
the lag direction, but the main contribution to the signal comes from
around 7~$\mu$s. The more normal lag centred at 22~$\mu$s is also
shown for comparison.
The range extents are rather similar
but the power is, of course, considerably lower for the pseudo zero lag.
Nevertheless, this is taken care of in the analysis and this lag is
rather important in events with high temperatures giving broad ion
line spectra or narrow ACFs. The transmitting frequencies were chosen to give
maximum radiated power for a given high voltage setting.

\begin{figure}[htbp]
  \begin{center}
    \includegraphics[width=\columnwidth]{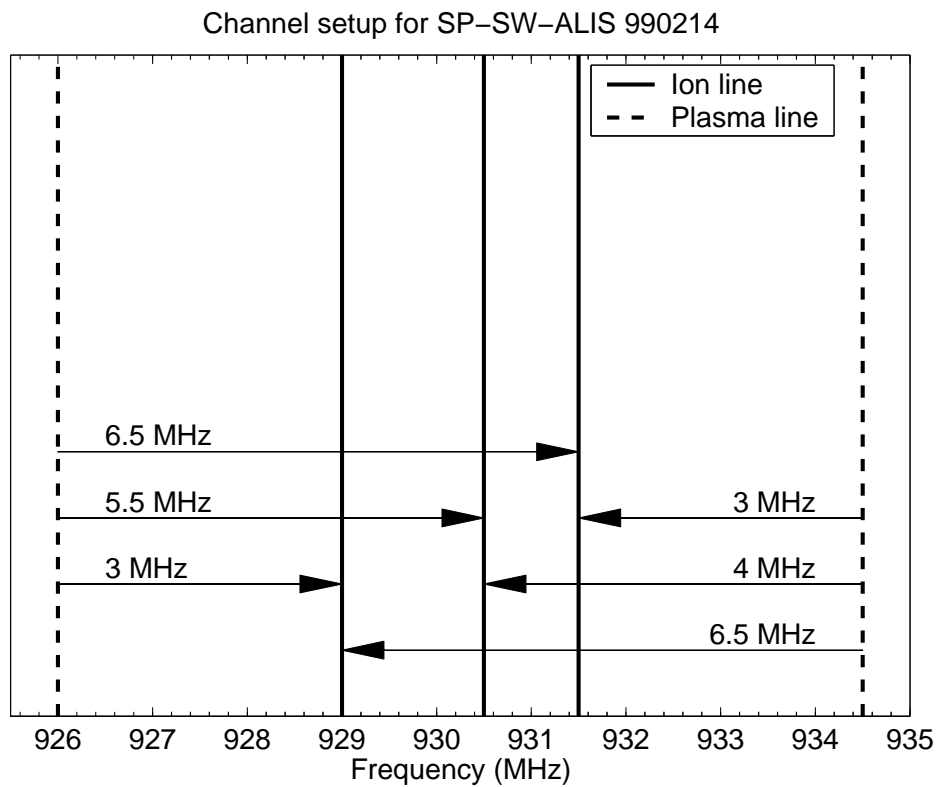}
    \caption{The frequency setup for the
      experiment. The plasma line channels were fixed to 926 and 934.5~MHz,
      giving "simultaneously" up and downshifted frequencies at 3
      and 6.5~MHz. In addition there is an upshifted line at 4 and a
      downshifted at 5.5~MHz.
    \label{fig:freq}}
  \end{center}
\end{figure}

The monostatic plasma line part of the experiment used two
channels covering the same ranges as the ion line but with 3.3~km
range separations between the spectra. The frequency setup for the experiment,
illustrated in \myfig{fig:freq}, gives the possibility for 3 upshifted
and 3 downshifted bands. For 3~MHz and 6.5~MHz frequency shifts, both
the up- and downshifted plasma lines were measured. In addition there
was a 4~MHz upshifted and a 5.5~MHz downshifted band. The width of
these bands should have been set to match the bit length of the codes
used, 50~kHz, but unfortunately this was not the case and 25~kHz wide
filters were erroneously used. This gave naturally less signal
throughput, and in \myfig{fig:amb} the corresponding range/lag ambiguity
functions for the plasma line channels are included for the first two
lags. The decoding still works, giving just slightly increased
unwanted ambiguities, but above all the pseudo zero lag is moved out
to a larger lag value. This fact, with one exception, almost ruled out
the possibilities to
measure plasma line spectra because, as will be shown here later,
they are likely to be rather wide, due to large time and height
gradients of the plasma parameters. This lead the analysis to use mainly the
undecoded zerolag, which after integration over the different codes,
almost resembles the shape of a 352~$\mu$s (16$\times22~\mu$s) long pulse.

The experiment contained a large
number of antenna pointings in order to follow the auroral arcs, but
as this day was cloudy over northern Scandinavia the transmitting antenna was
kept fixed along the local geomagnetic field line. The remote sites,
receiving only ion lines,
were monitoring the same pulses as the transmitting site, and were
used to measure the drifts in the F-region, to derive the electric
field. Thus, these antennas intersected the transmitted beam at
the F-region altitude giving the best signal, for this day mostly at
170~km.

\section{Measurements}
\label{sec:measurements}

\subsection{Ion line}
\label{sec:ion-line}

\begin{figure*}[htbp]
  \begin{center}
    \includegraphics[width=\textwidth]{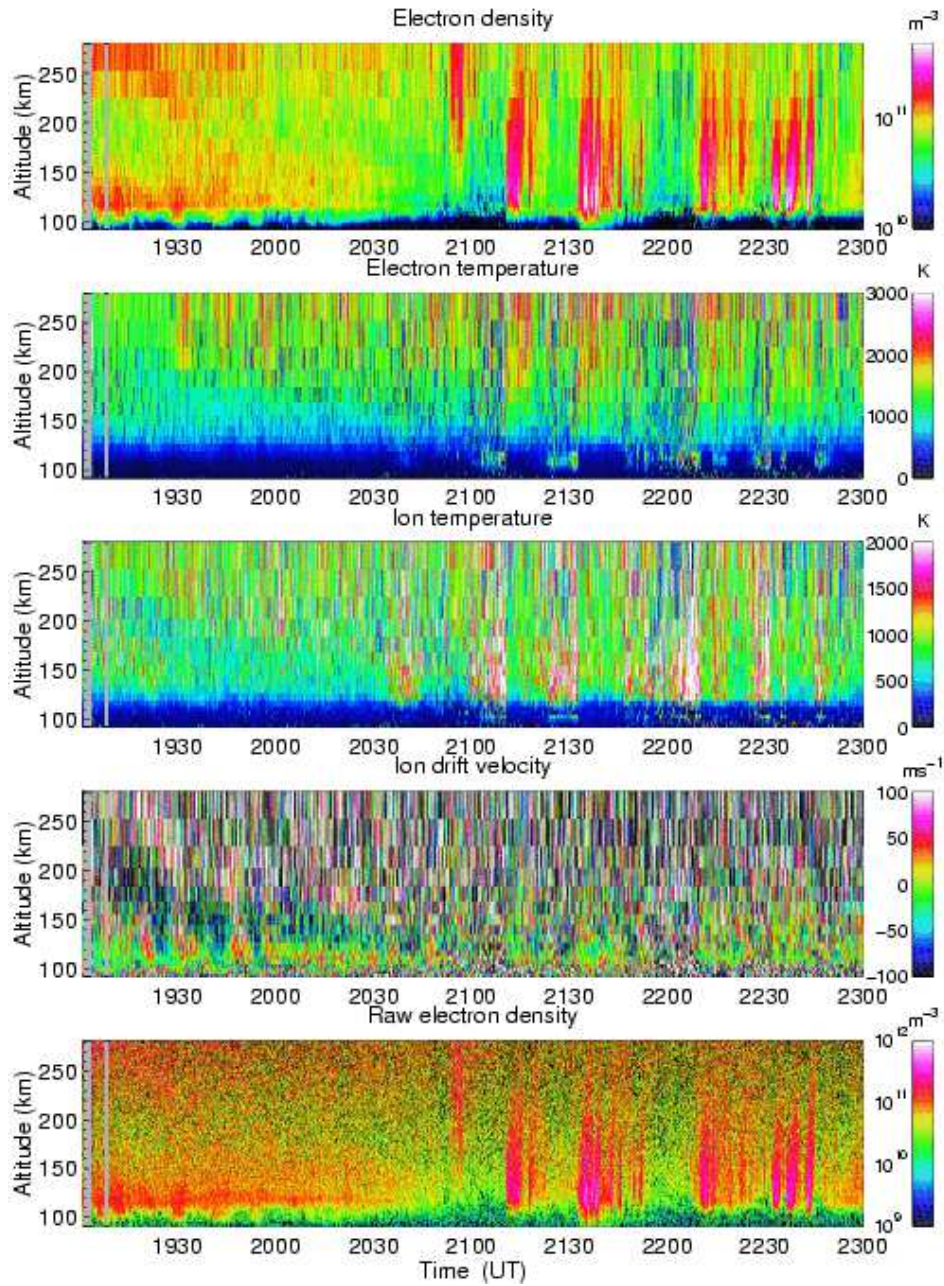}
    \caption{Summary of results from the ion line experiment. All
      panels shows the parameters in a altitude versus time fashion
      with the antenna directed along the geomagnetic field line.
      Panels from top: Electron density, electron temperature, ion
      temperature, line-of-sight ion drift velocity and electron
      density based on only the returned power. The data were analysed
      with 5~s integration due to the active conditions.
    \label{fig:ionover}}
  \end{center}
\end{figure*}

The experiment started at 1900~UT on 14 February 1999 and continued until
2300~UT. \myFig{fig:ionover} shows an overview of the parameters deduced
from the ion line measurements, which were analysed using the on-line
integration time, 5 seconds, in order to be compared to the plasma
line data. The analysis was done using the GUISDAP package
\citep{lehtinen1996jatp}, but a correction of 45\% of the radar system
constant used in the package had to be invoked to fit the
plasma line measurements according to \myeq{eq:1}.
This short integration time was possible due to the highly
optimised mode used, with all the transmitter power concentrated to
the E-region. In range, some integration was done, so that at lower
heights 2 range gates were added together and with increasing height
the number of gates added together increases to 15 in the F-region.

There is a rather strong E-region from the start,
but no real arcs, and we interpret this as diffuse aurora. The peak
electron density shows some variation, but as time goes the
E-region ionisation decreases until 2050~UT, where it is almost gone.
The density
peak during this time was at around 120~km altitude and the lower edge
of the E-region
at 110~km, but at times the ionisation reaches down to 100~km. At 2050~UT
and onwards until 2240~UT the ionosphere above Tromsø became more
active and several auroral arcs passed the beam. Around some of the
arcs there are short-lived enhancements of electric field, seen as
F-region ion and E-region electron temperature increases. In the last
10 minutes of the experiment the arc activity
disappeared and again there was diffuse aurora. From the field aligned
ion drifts it is evident that there is a rather strong wave activity in the
diffuse aurora until 2050~UT, while it is not so clear in the
continuation of the experiment. The last panel with the inferred
electron density from the pseudo zero lag shows the same features as
the fitted density panel but with highest possible resolution since no
height integration is made.

\subsection{Plasma line}
\label{sec:plasma-line}

\begin{figure}[htbp]
  \begin{center}
    \includegraphics[width=\columnwidth]{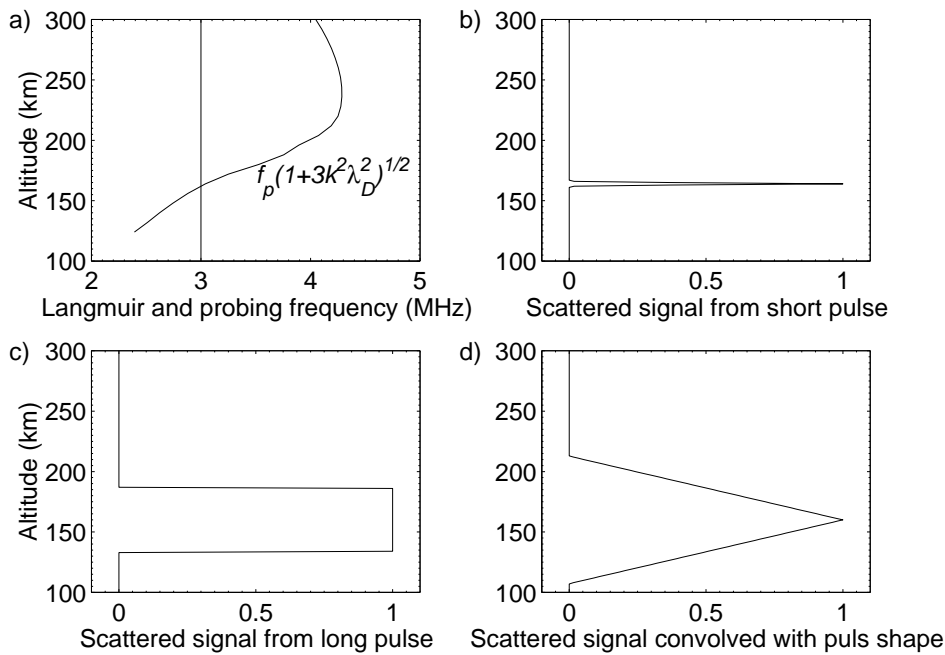}
       \caption{Description of the analysis
         procedure. a) Langmuir frequency profile together with the
         probing frequency of 3~MHz shift. b) The echo profile using a
         short pulse. c) The echo profile using a long pulse. d) The
         echo profile of a long pulse convolved with the pulse shape.
    \label{fig:planal}}
  \end{center}
\end{figure}

Since the analysis of the plasma lines was forced to handle the
undecoded zerolags of the alternating codes, it was necessary to
analyse their profile shape. \myFig{fig:planal} shows how this analysis
was performed. From the fitted parameters of the ion line, electron
density and temperature, a profile of the approximate Langmuir
frequency can be calculated using \myeq{eq:1}.
When probing at fixed frequency, there
will be scattered signal only from heights where the probing and
Langmuir frequencies match each other. The effect of the undecoded
zerolag is similar to the one where a normal long pulse is used, but
with a lower signal strength. So, the profile shape will be a
square pulse centred on the corresponding altitude, since no gating is
performed. Because the signal
strengths are rather weak compared to the system temperature, there is
a great deal of noise in the profile shape. To be able to
extract the altitude and signal power, a fitting procedure need to be
performed. For this a continous function and a good first guess is
needed and the measured plasma line profile was convolved with the
pulse shape to have a triangular shape and also showing a
peak close to the matching height.

\begin{figure*}[htbp]
  \begin{center}
    \includegraphics[width=\textwidth]{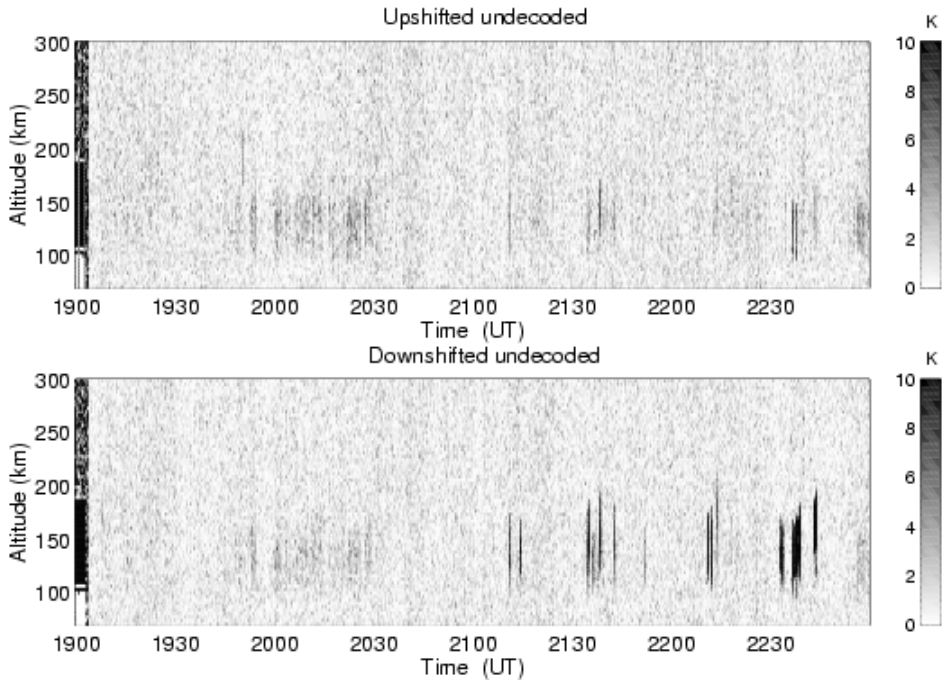}
    \caption{Overview of the plasma line measurements. The panels
      show the undecoded upshifted (top panel) and downshifted
      returns, given as signal strengths in Kelvin.
    \label{fig:plover}}
  \end{center}
\end{figure*}

The plasma line part of the experiment is overviewed in \myfig{fig:plover}.
The signal strengths shown should be compared with the UHF system
temperature of about 90~K. At first glance there is almost nothing in
the upshifted part, but a more careful look shows weak signals between
1940 and 2030~UT and after 2250~UT, corresponding to the occurrence of
diffuse aurora. Similar echoes can also be seen in the downshifted
channel, and are due to plasma lines at 3~MHz offset from the
transmitted frequencies, according to the ion line measurements.
In the downshifted part after 2100~UT,
frequent events of rather strong signals in phase with auroral arcs
pass the beam. Most of these are from the 5.5~MHz shift, but some
of them also are due to plasma lines at 6.5~MHz. Such events are less
frequent and for most of them there are also signals in the upshifted
part.

\begin{figure}[htbp]
  \begin{center}
    \includegraphics[width=\columnwidth]{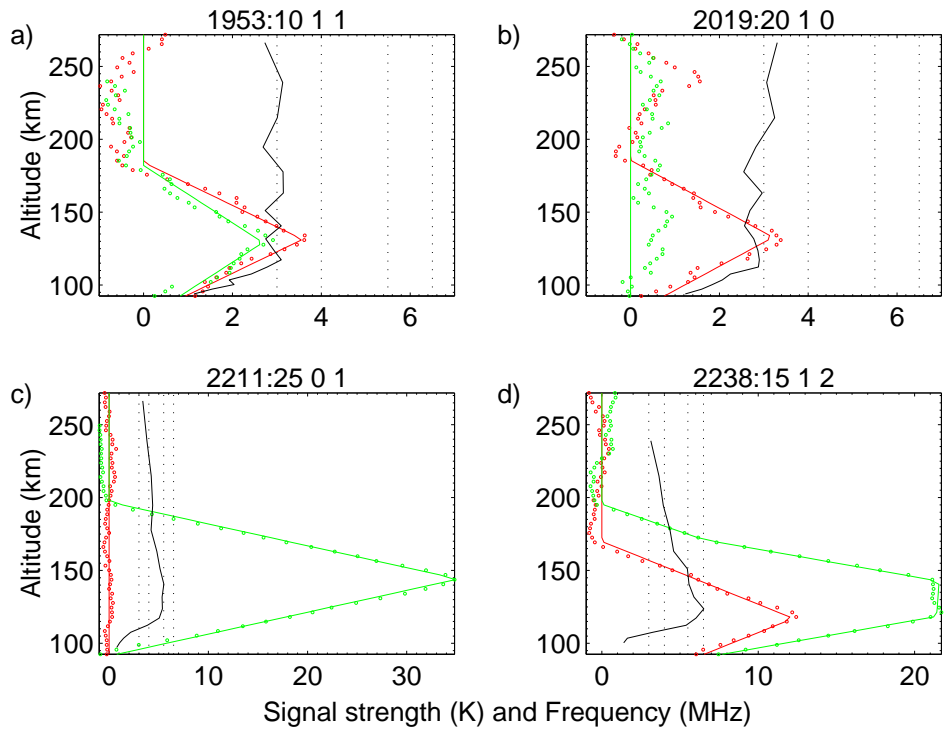}
          \caption{Examples of measured convolved profiles (circles) and the 
            triangle(s) fitted (lines).
            Included on the plots are the profiles of the
            Langmuir frequency (black) calculated from the parameters
            of ion lines fit. The value on the x-axis represent signal
            strength (K) and frequency (MHz) respectively. The numbers
            in the title shows the time of measurement and how many
            echoes was detected for the up- (red) and downshifted
            (green) frequencies. A lower limit of 2~K received signal
            power (2\%~SNR) was used. a) Simultaneous up and
            downshifted echoes at 3~MHz in diffuse aurora. b) An
            upshifted 3~MHz echo in diffuse aurora. c) A strong
            downshifted echo at 5.5~MHz in an auroral arc. d)
            Simultaneous up and downshifted echoes at 6.5~MHz and a
            downshifted 5.5 in an auroral arc.
    \label{fig:plex}}
  \end{center}
\end{figure}

Due to the fact that the same channel is used for several frequencies,
there can be several altitudes that fulfill the matching condition
between Langmuir and probing frequency. This complicates the analysis
somewhat, and there have to be several fits with different numbers of
triangles superposed on each other. Some examples of this analysis are
shown in \myfig{fig:plex}, from only one plasma line to several both up-
and downshifted lines. Using a lower limit of 2~K signal power, the total
number of 5 second integration events with enhanced plasma lines for
this evening
was 256, and a total of 468 plasma line echoes were detected, divided
into 220 on 3~MHz, 19 on 4~MHz, 157 on 5.5~MHz and 72 on 6.5~MHz.

\section{Theory}
\label{sec:theory}

\begin{figure}[htbp]
  \begin{center}
    \includegraphics[width=\columnwidth]{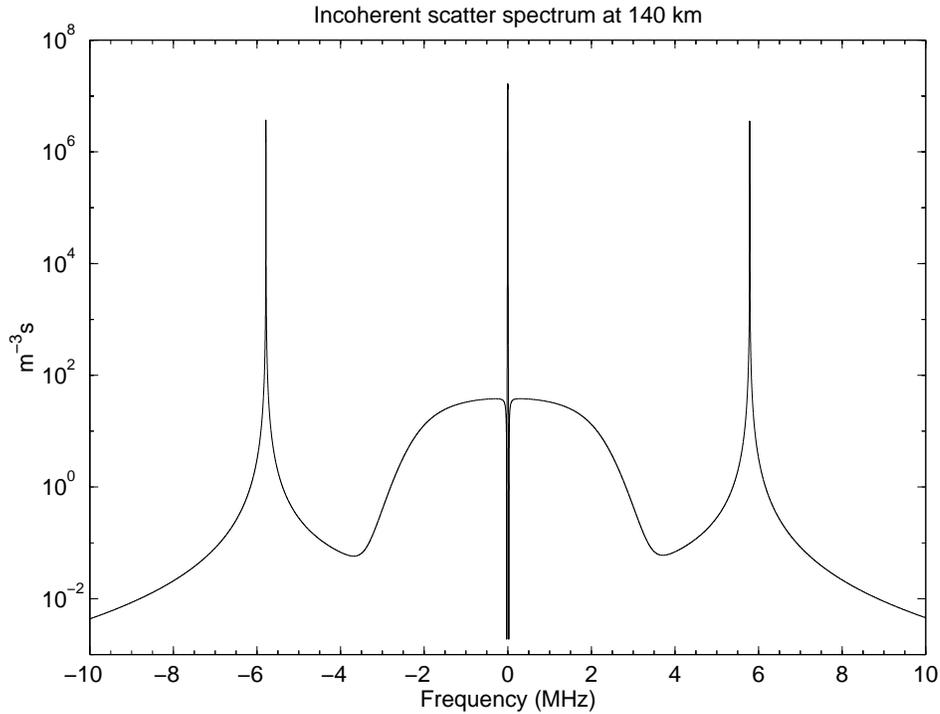}
    \caption{The wide frequency incoherent scatter spectrum. The parameters
      used: $N_e=4\cdot10^{11}$~m$^{-3}$, $T_e=700$~K, $T_i=600$~K,
      $v_i=v_e=0$~ms$^{-1}$ and collision frequencies $\nu_e$ and
      $\nu_i$ using the MSIS90e model (Hedin, 1991) for 140~km
      altitude. Logarithmic scale for the strength is used to be able
      to emphasise the different lines of the spectrum.
    \label{fig:wideis}}
  \end{center}
\end{figure}

In order to relate the plasma line measurements to physical quantities
it is necessary to investigate the spectrum of the incoherent scatter
process. The Nyquist theorem approach, derived in a long series of papers
by \citet{dougherty1960procrsl,dougherty1963jgr}, \citet{farley1961procrsl},
\citet{farley1966jgr} and finally \citet{swartz1979jgr}, arrives at
\begin{equation}
  \label{eq:2}
  \sigma(\omega) =\frac{N_e r_e^2 \sin^2\delta} {\pi} \cdot 
  \frac
  {\left| 
      y_e 
    \right|^2 
    \sum_i \frac{ \eta_i \Re (y_i) } {\omega-\vec k \cdot \vec v_i} +
    \left| 
      jk^2\lambda_D^2 + \sum_i \mu_i  y_i 
    \right|^2 \frac 
    { \Re (y_e) } {\omega-\vec k \cdot \vec v_e}} 
  {\left| 
      y_e  + jk^2\lambda_D^2 + \sum_i \mu_i y_i  
    \right|^2},
\end{equation}
where
\begin{equation}
  \label{eq:3}
  \eta_i=\frac{n_i q_i^2}{N_e e^2},
\end{equation}
\begin{equation}
  \label{eq:4}
  \mu_i=\frac{\eta_i T_e}{T_i},
\end{equation}
and the index $e$ stands for electrons and $i$ for the different ion species.
$N$ and $n$ are the densities, $r_e$ the classical electron radius, $v$ the bulk velocity, $q$ the
charge, $e$ the electron charge and $T$ the temperature.
The complex normalised admittance function, $y$, contains most of the
physics with the plasma dispersion function and have as main arguments
the collision frequency and magnetic field. 
The same result
was also reached by \citet{rosenbluth1962pf}, using the dressed particle
approach. In \myfig{fig:wideis} there is an example of the spectrum,
showing clearly the strong ion line around zero offset frequency and
the rather weak plasma lines at rather large offsets. The plasma lines
become enhanced by a photo electron or auroral electron produced
suprathermal electron distribution, and in order to simulate what this
extra distribution does to the spectrum a modification to the formula has
to be made. First, a rewriting of \myeq{eq:2} following
\citet{swartz1978rs} has to be performed
in order to separate the electron and ion contributions to the
spectrum:
\begin{equation}
  \label{eq:5}
   S(f) = \frac{1} {\pi} \cdot 
   \frac{ \left| \frac{N_e y_e}{T_e} \right|^2
    \sum_i \frac{ n_i Z_i^2 \Re(y_i) } {f+k v_i/2\pi} +
    \left|  jC_D + \sum_i \frac {n_i Z_i^2 y_i}{T_i} \right|^2 \frac {
      N_e \Re(y_e) } {f+k v_e/2\pi}} 
  {\left| \frac{N_e y_e}{T_e}  + jC_D +
    \sum_i \frac{n_i Z_i^2 y_i}{T_i} \right|^2},
\end{equation}
where
\begin{equation}
  \label{eqcd}
  C_D=\frac{k^2 \epsilon K}{e^2},
\end{equation}
$f$ is the frequency shift, $Z=q/e$, $\epsilon$ is the dielectricity
and $K$ is the Boltzmann constant. Here, a
normalisation of the spectrum has also been performed, so that the zero lag
of the corresponding ACF is the raw electron density and the vector
velocity is replaced by the line-of-sight velocity. Using a treatment
in analogy to the ion contribution, it is now possible to rewrite the
spectrum to support a number of Maxwellian electron distributions as
\begin{equation}
  \label{eq:6}
     S(f) = \frac{1}{\pi} \cdot 
   \frac{ \left| \sum_e \frac{N_e y_e}{T_e} \right|^2
    \sum_i \frac{ n_i Z_i^2 \Re(y_i) } {f+k v_i/2\pi} +
    \left|  jC_D + \sum_i \frac {n_i Z_i^2 y_i}{T_i} \right|^2 \sum_e \frac {
      N_e \Re(y_e) } {f+k v_e/2\pi}} 
  {\left| \sum_e \frac{N_e y_e}{T_e}  + jC_D +
    \sum_i \frac{n_i Z_i^2 y_i}{T_i} \right|^2}.
\end{equation}

\begin{figure}[htbp]
  \begin{center}
    \includegraphics[width=\columnwidth]{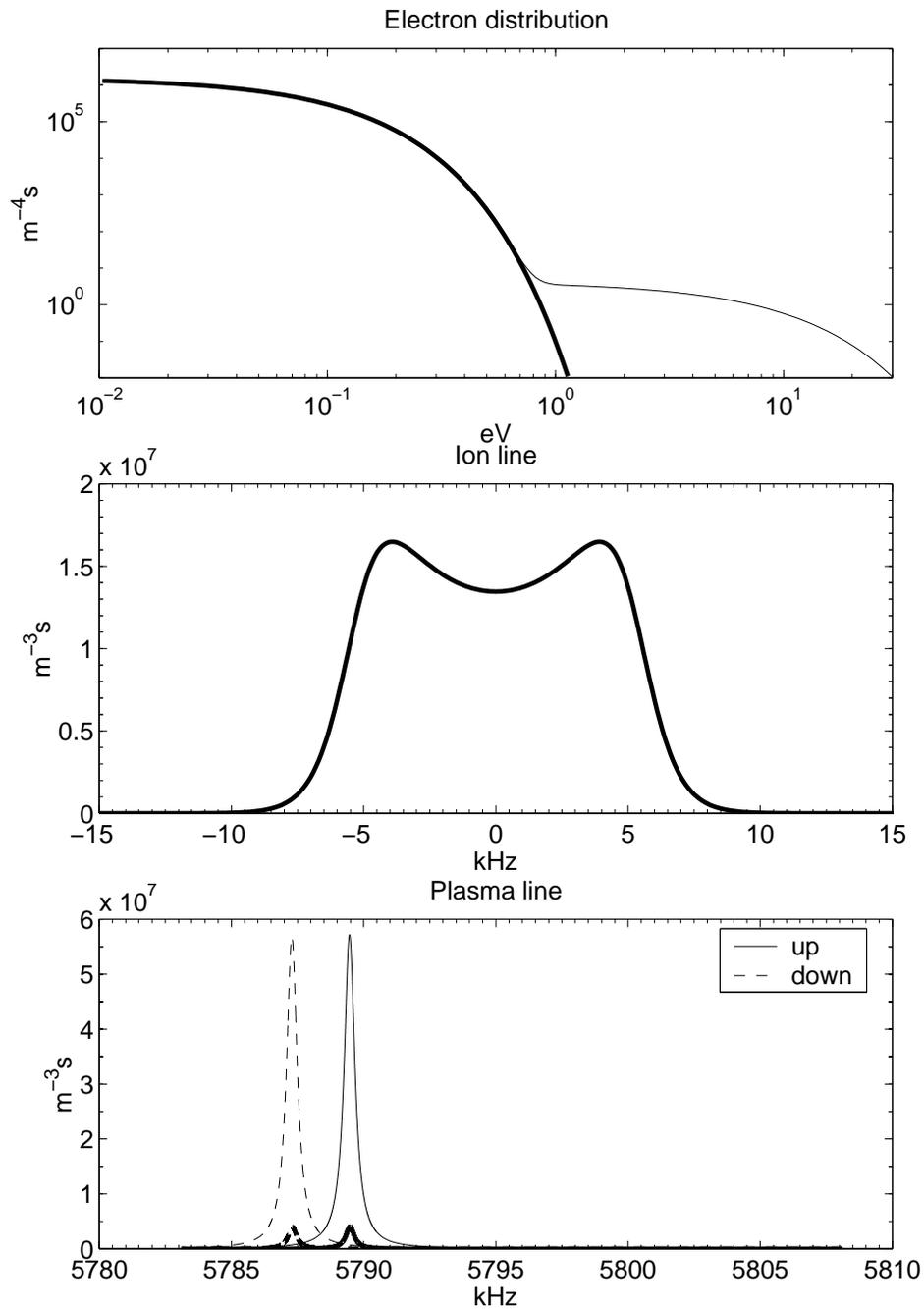}
    \caption{Close-up of the ion and the plasma lines. Top panel
      shows the electron distribution, middle panel the ion line and
      bottom panel the plasma lines, where the frequency scale of the
      downshifted line have been reversed. The figure shows the lines
      with two different electron distributions, thick line for the
      normal thermal distribution and thin line for the same
      distribution together with an suprathermal distribution.
    \label{fig:narrowis}}
  \end{center}
\end{figure}

\begin{figure}[htbp]
  \begin{center}
    \includegraphics[width=\columnwidth]{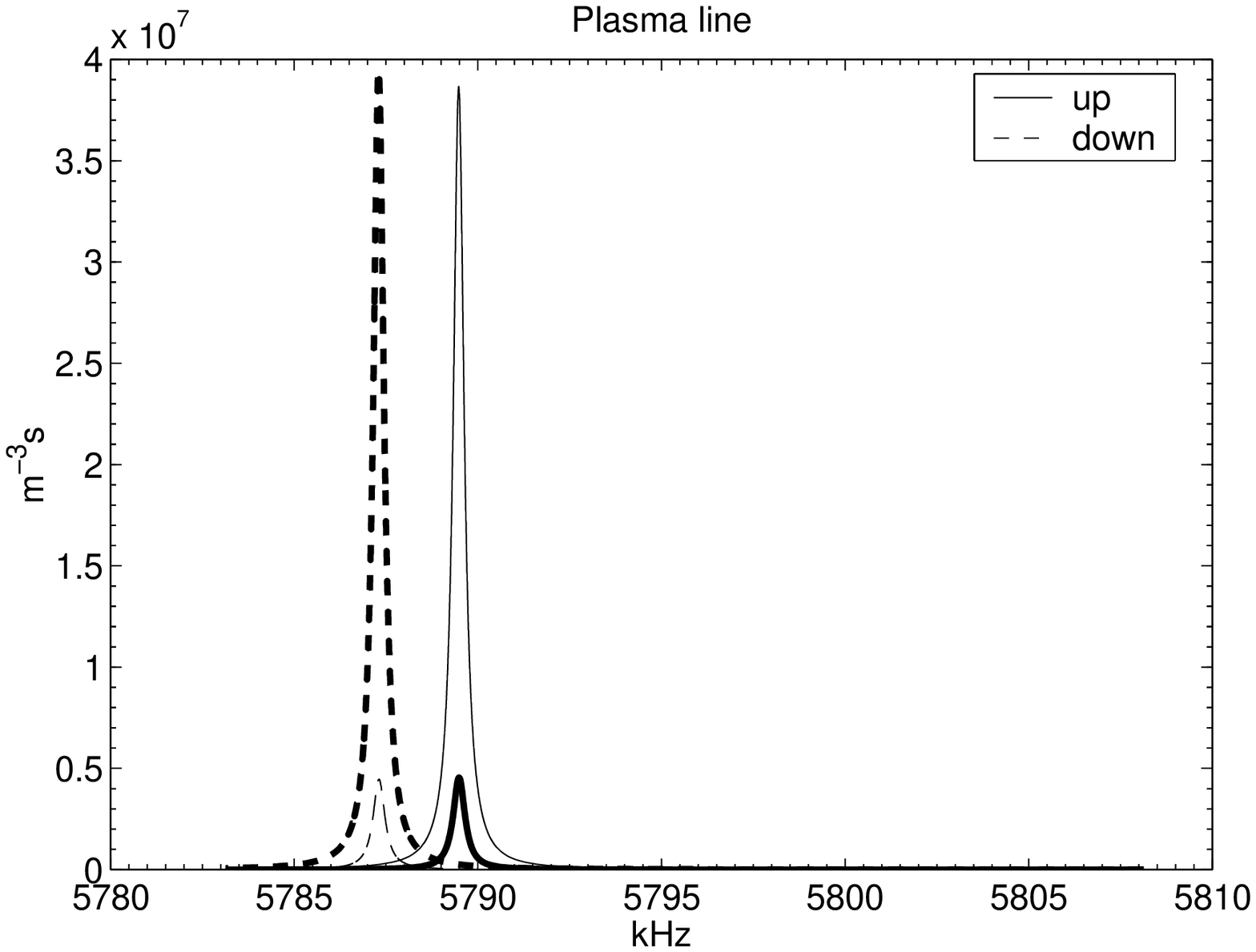}
       \caption{The plasma lines based on an upgoing (thick line)
         and downgoing 10~eV beam of suprathermal electrons at a
         current of 1 $\mu$Am$^{-2}$ and temperature of 5~eV. The
         thermal parameters are the same as in \myfig{fig:wideis}
    \label{fig:plstrength}}
  \end{center}
\end{figure}

In \myfig{fig:narrowis} the effect on the plasma lines of a suprathermal
distribution with a reasonable density of $10^7$~m$^{-3}$ and width of 10~eV
is shown, being lower than the ionisation energy for most ions. The
plasma lines grow considerable and the integrated power over the
bandwidth used in the experiment becomes comparable to the power of
the ion line. It may also be noted that there is no effect at all seen
in the ion line. \citet{perkins1965pr} have shown similar
calculations, but when their method was based on large expansions to
allow non-Maxwellian distributions one can here more directly superpose a
few Maxwellian distributions to explain the measurements and even make
fits of the spectra taken to get estimates of the suprathermal
distributions, which is of great importance in auroral measurements.
A consequence of \myeq{eq:6} is that it is also possible to
derive the spectrum assuming currents carried by the suprathermal
distribution, since it allows different drift velocities on the
various distributions. Indeed, \myfig{fig:plstrength}, shows differential
strengths on the two plasma lines, with upgoing electrons enhancing mainly the
downshifted line and downgoing ones the upshifted line.

\section{Discussion}
\label{sec:discussion}

Plasma line measurements in the active auroral ionosphere are not an
easy task, due to the large variations in the ionospheric parameters.
The Langmuir frequencies are largely dependent on the ambient electron
density, making the line move considerably as the density changes,
which it does on time scales of seconds. Moreover, the density height
gradient makes the lines very broad when measured over a specific
height interval, and at times even broader than the receiver
band. A chirped radar would solve only a part of the problem at the
cost of transmitter power.
These complications make it hard to draw any conclusions on the power
in the lines, as one does not know for sure the scattering volume or the
time duration of the scattering. Bearing this in mind and to at least
minimise these effects, one can nevertheless look at the different
distributions in height and power for the different lines to get an
idea of their nature, using the on-line integration time of 5 seconds.

\begin{figure}[htbp]
  \begin{center}
    \includegraphics[angle=90,width=\columnwidth]{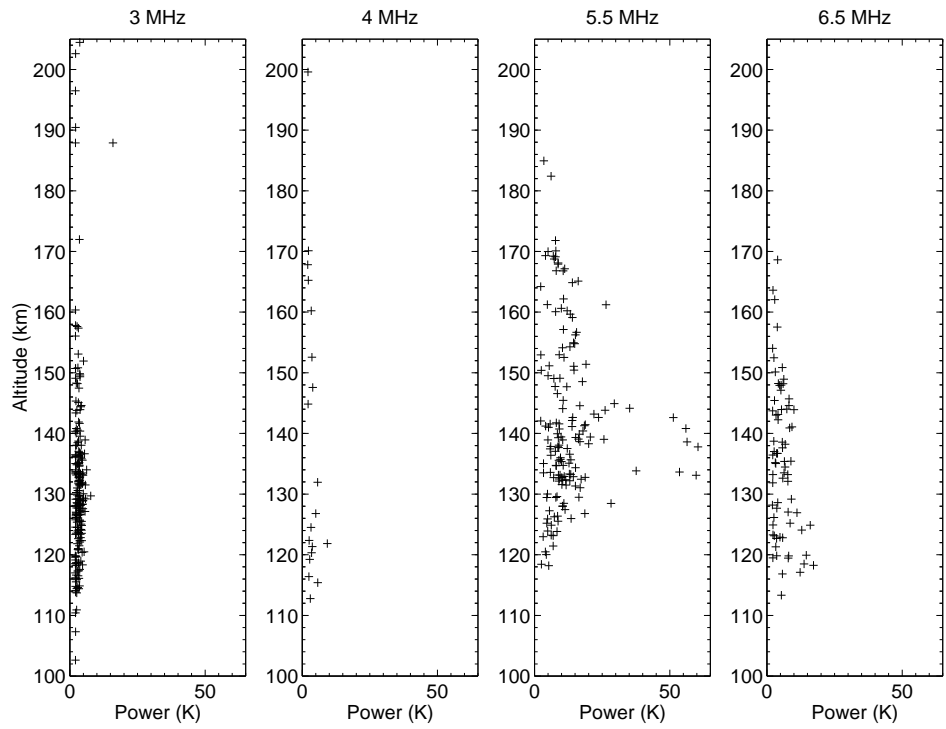}
    \caption{Scattered plots of echo
      altitude and strength for the different frequency shifts. The
      altitudes range from 100 to 210~km, E and lower F region.
    \label{fig:scatterall}}
  \end{center}
\end{figure}

\begin{figure}[htbp]
  \begin{center}
    \includegraphics[width=\columnwidth]{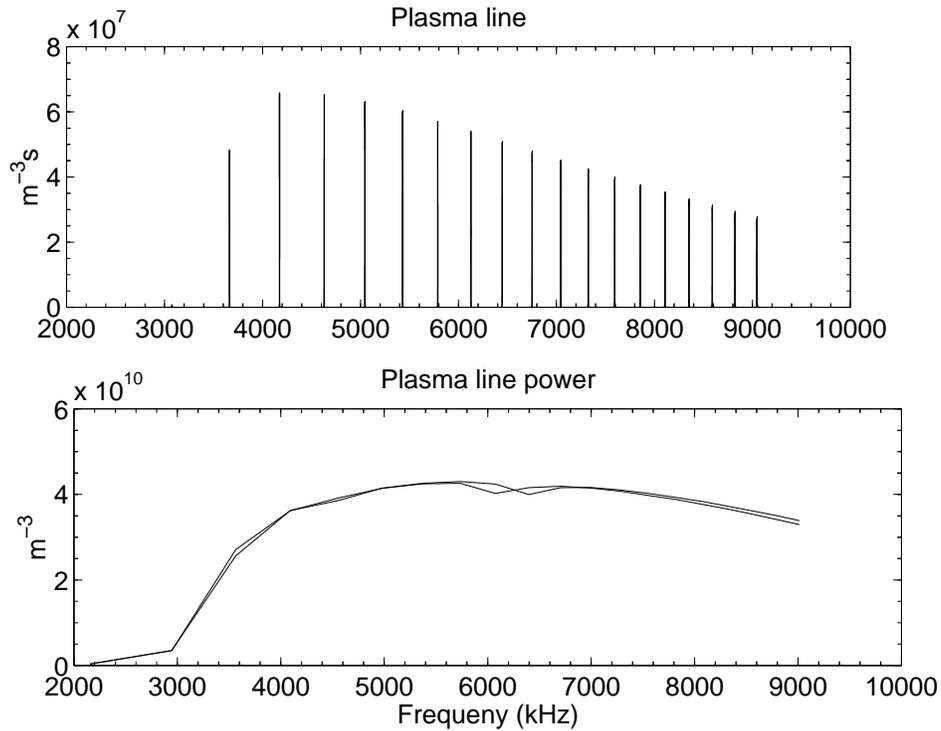}
    \caption{Simulation of the plasma line strength at 140~km altitude with an
      added distribution of secondaries for different electron
      densities from 0.5 to $10 \cdot 10^{11}$ m$^{-3}$. The envelope
      in the lowest panel shows the total power of the plasma lines
      versus the Langmuir frequency.
    \label{fig:simelpower}}
  \end{center}
\end{figure}

The altitude distributions for the different frequencies are shown in
\myfig{fig:scatterall}. One must note that the signal levels shown are not
corrected for range, as the scattering volumes are not known, so
signals from a higher altitude are in fact stronger than the
corresponding signal from a lower height. It is clearly seen that the
strongest signal is the 5.5~MHz line, followed by 6.5, 4 and 3~MHz.
The altitude distribution shows more or less the expected dependence
on range, but there are some exceptions: In one point at 188~km in the
3~MHz band and for the 5.5~MHz band the strong values between 130 and
140~km seem to be stronger than the others, even taking into account
the range effect. However, the number of points are too few to be used
as evidence on altitude effects. Most of the echoes are coming from around 
120-150~km altitude and \myfig{fig:simelpower} shows a simulation of the
expected strength of the plasma line for given background electron
density. It shows a peak at around 5.5~MHz and this is also what the
experiment shows. A more realistic suprathermal distribution will
decrease the returned power for a number of frequencies and one should
see the figure as an upper-limit estimate. Indeed, \citet{nilsson1996anngeo},
have made predictions of the expected strength of Langmuir
waves for different heights and carefully derived distributions. These
predictions are in rather good agreement with the present measurements
showing a strong peak between 5 and 6.5~MHz.
Although there is some uncertainty on the scattering volume, the
observed strengths of the plasma lines can be used to get some
estimates of the distribution of the suprathermal population. 
Assuming a width of 6~eV, one need, to get to the measured
strengths for the 3~MHz case, a suprathermal density of about
$5\cdot10^8$~m$^{-3}$ which is 0.5~\% of the thermal population.
For the 5.5 and 6.5~MHz cases, it is enough with only
$1\cdot10^8$~m$^{-3}$, but the uncertainty of volume is even larger
here due to the active environment and the values should maybe be the same as
for the lower frequency offsets.

\begin{figure}[htbp]
  \begin{center}
    \includegraphics[angle=90,width=\columnwidth]{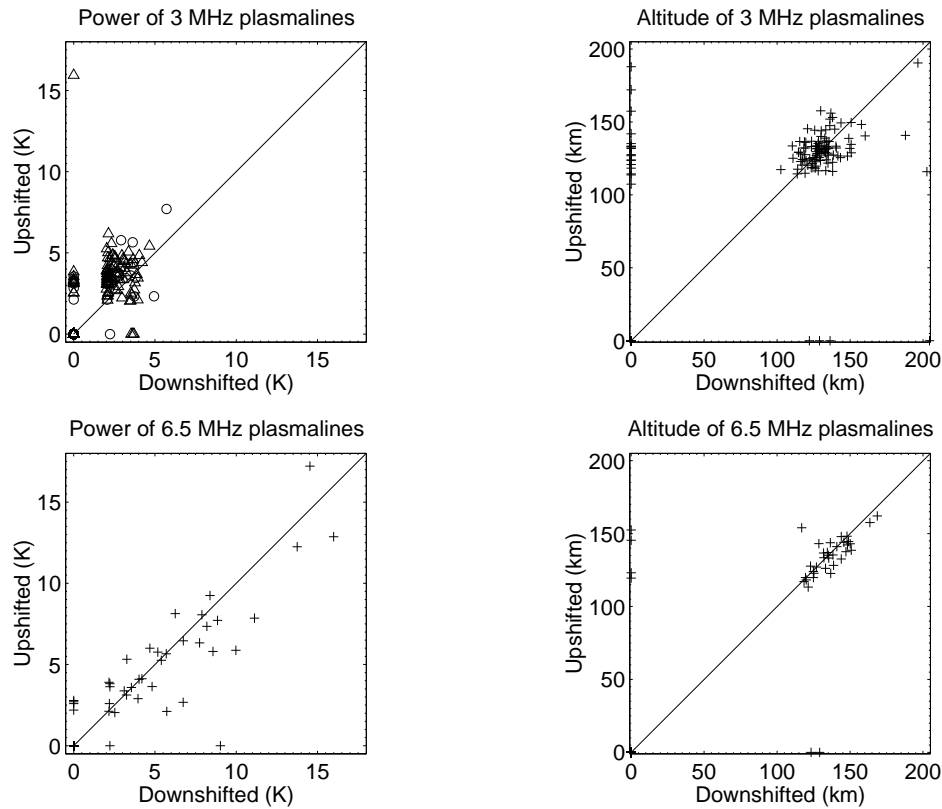}
    \caption{Scatter plots of power and altitude for the two simultaneously
      measured up and downshifted lines at 3 and 6.5~MHz. The
      triangles and circles for the power of the 3~MHz case, in the
      upper left, show lines detected before and after magnetic
      midnight respectively --- no difference can be seen.
    \label{fig:scatter36}}
  \end{center}
\end{figure}

\begin{figure}[htbp]
  \begin{center}
    \includegraphics[width=\columnwidth]{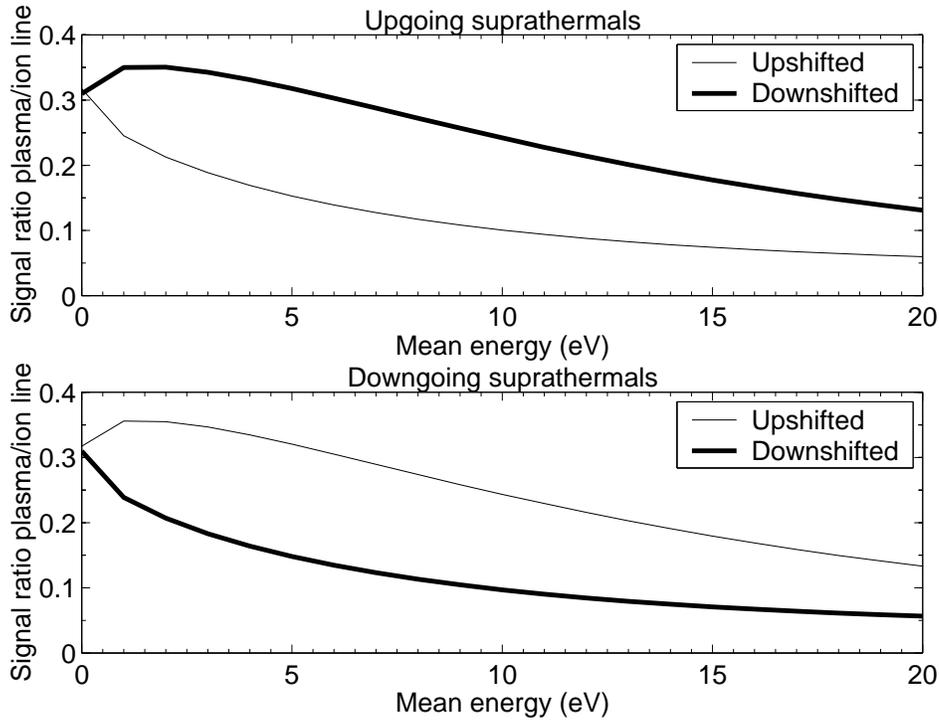}
    \caption{Simulation of 3~MHz plasma line strength
      at 140 km with an electron beam with varying mean energy from 0
      to 20~eV and temperature of 10~eV, corresponding to a few
      $\mu$Am$^{-2}$ field-aligned currents, depending on energy. The top
      panel shows the ratio between plasma and ion line strengths for
      upgoing beams and the bottom for downgoing beams.
    \label{fig:simfreqpower}}
  \end{center}
\end{figure}

The most interesting
thing with incoherent plasma lines is, of course, the possibility to
derive differential drifts between ions and electrons, and from these
deduce ionospheric currents. For this, one needs to measure the up-
and downshifted lines simultaneously. Although the time of the
measurement for both lines were not exactly the same in this
experiment, the time shifts between them are so small (3-6~ms)
compared to the total cycle time (300~ms) of the codes, that this
effect is of minor importance. In \myfig{fig:scatter36}, the strengths and
altitudes of the two concurrently recorded up- and downshifted plasma
lines are shown. In general, there are stronger up- than downshifted
lines for the 3~MHz case, whereas no such trend can be seen in the 6.5
MHz band. However, there are exceptions to these overall trends
and on occasions there are large differences in the signal strengths
between the lines. Almost all of the 3~MHz plasma lines were recorded
in diffuse aurora and this evident difference in signal power needs a
closer examination. When there is a drift of the thermal electrons, the
plasma lines shift, and when probing at a fixed frequency, the scatter
may not come from the same altitude for the up- and downshifted lines
respectively. The strength of the plasma line is also rather altitude
dependent due to the damping by the collisions of electrons with ions and
neutrals. But \myfig{fig:scatter36} shows no general height difference
between the up- and downshifted 3~MHz lines, so this difference in strength
cannot be explained by thermal electron bulk drifts. To simulate the effect of
current carried by suprathermal electrons, \myfig{fig:simfreqpower}
illustrates the strength of the 3~MHz plasma lines for Maxwellian
electron beams of different energies. With no net current the lines
are of almost the same strength, and the difference is mainly dependent on
where in the receiver band the lines are. However the experiment shows
stronger upshifted lines, thus it is evident that the diffuse aurora this
night contained fluxes of downgoing suprathermal electrons or, in other
words, there was an upgoing current carried by suprathermal electrons.
The average power ratio between up- and downshifted lines, 1.4, can be
used to estimate the amount of current carried by the suprathermals.
As before, using a density of $5\cdot10^8$~m$^{-3}$ and temperature of 6~eV,
and shifting this population to a current
density of 15~$\mu$Am$^{-2}$ brings to the observed ratios. This
current seems somewhat high, but not unrealistic.

For the 6.5~MHz bands there may be a slight difference with respect to
the altitude, and that is most likely due to thermal currents causing
frequency shifts of the plasma lines. This effect is not very clear,
but as lower heights have higher density, or Langmuir frequency, and
as the downshifted line is at a slightly lower height, this is most
probably an effect of a downgoing current carried by thermal
electrons. The correction due to heat-flow in the plasma dispersion
function, discussed by \citet{kofman1993jgr} and confirmed later also
by \citet{nilsson1996jatp} and \citet{guio1996anngeo}, but not taken into
account here, would also show the same effect in altitude difference
between the lines. The band widths used here, 25~kHz, are much wider
than the effect of heat-flow, which is less than or around 1~kHz in the
F-region and much lower in the E-region, so that cannot explain the
6.5~MHz height differences. A recent paper by \citet{guio1998anngeo}
with proper calculation of the dispersion equation 
investigates the heat-flow and finds that it is not necessary
to invoke the effect at all, but as this paper don't go to the same
extreme the heat-flow has to be considered.

\begin{figure}[htbp]
  \begin{center}
    \includegraphics[width=\columnwidth]{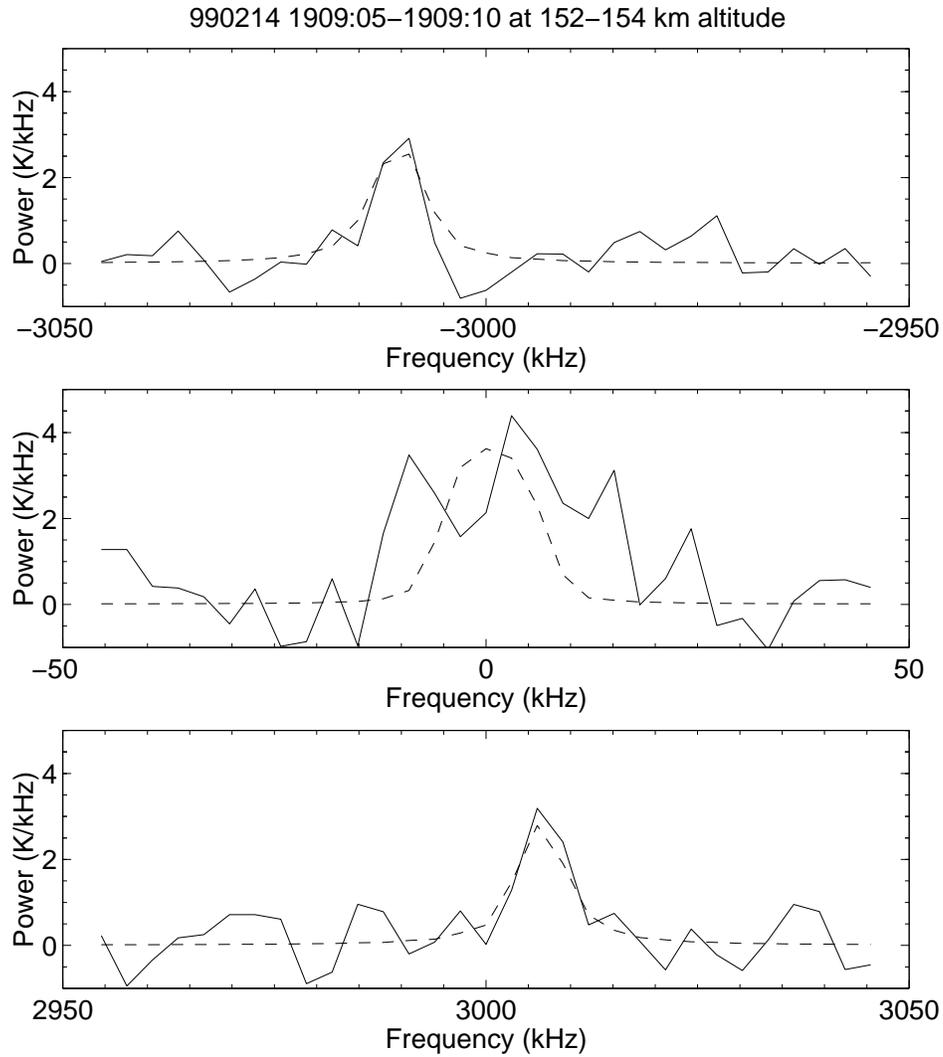}
    \caption{Measured spectra of the both plasma
      lines and the ion line and the best 7-parameter fit of the
      theoretical spectrum. The parameters fitted were $N_e=9.7 \cdot
      10^{10}$~m$^{-3}$, $T_e=631$~K, $T_i=697 $~K,
      $v_i=-100$~ms$^{-1}$,
      $v_e=657$~ms$^{-1}$ and a suprathermal distribution
      with $n_e=0.8 \cdot 10^9$~m$^{-3}$ and $T=11$~eV.
    \label{fig:fullfit}}
  \end{center}
\end{figure}

Most of the above discussions on currents are only on directions,
but to get any quantitative numbers it is necessary to 
look at the individual spectra. Of the 256 events of enhanced plasma
lines, there is only one that is good enough to investigate. All the
other are too broad either due to the Langmuir frequency gradient
smearing or the fact that the Langmuir frequency at the height in
question is varying
during the 5 second time slots. A further reason is that the
stationarity condition for the alternating code technique is not
fulfilled, due to changes of Langmuir frequency, and the spectra
change too much within the 300~ms cycle. Anyway, there is one, and
the up- and downshifted bands, together with the ion line band, are
shown in \myfig{fig:fullfit}. The shifts in this case are around 3~MHz as
deduced from the ion line analysis. These spectra were then fitted to
the theoretical spectrum in \myeq{eq:6} using 7 ionospheric
parameters, the thermal parameters for both ions and electrons
and the parameters of a non-shifted suprathermal electron distribution.
At the first glance the fit on the ion line seems rather poor, but the
fit was done as usual in the time domain with proper weighting on the
different lags of the ACF, and in the FFT process to produce
\myfig{fig:fullfit} these statistical properties are lost. The fit
looks actually better in the time domain, but the frequency domain was
chosen in the figure to be more informative.
The current density $j=N_ee(v_i-v_e)$, and with the fitted parameters, the
field aligned current carried by the thermal electrons, amount to
12~$\mu$Am$^{-2}$ downward. This is opposite the general current seen in
the plasma line strength for the 3~MHz band, but the magnitudes are
comparable. For this case no current could be seen on the
suprathermals, and it is evident that the currents are rather
structured. Of course, there are many more cases of no plasma lines
than enhanced plasma lines in the diffuse aurora, and some cases of
only one plasma line either up- or downshifted. It is not surprising that
this single example of thermal currents do not follow the general trend.
Again, heat-flow is not
taken into account, but that should increase the current somewhat.
The deduced suprathermal distribution is in good agreement with those
\citet{kirkwood1995jgr} derived from the precipitating flux of
primaries; of course in the present case the distribution is
Maxwellian and without the fine structure due to the atmospheric
constituents, but the numbers are comparable.

\section{Conclusions}
\label{sec:conclusions}

We have made measurements of plasma lines in the active auroral
E-region. During 256 periods of 5 second integration we found a total
of 468 plasma line echoes, divided into 220 on 3~MHz, 19 on 4~MHz, 157
on 5.5~MHz and 72 on 6.5~MHz. It may seem strange to try to measure
plasma lines at such a low frequency as 3~MHz giving low signal
levels, but in fact most of the echoes and the most interesting results
came from this frequency offset.

The strongest echoes were found at the 5.5~MHz line, and somewhat
weaker ones at 6.5~MHz, inside auroral arcs. One must, however, note
that the strength measured inside the arcs is mostly a low-limit
estimate due to the active environment. The integration period used, 5~s,
is rather long in auroral arc conditions, and changes typically occur on
shorter time scales. Therefore, effects of gradients in the
Langmuir frequency profile, and hence scattering volumes, have not
been taken into account.

The Holy Grail in incoherent scatter plasma lines is the possibility
to measure currents, and in this case, the field-aligned currents in
aurora. The simulations carried out here, the extended full incoherent
scatter spectrum with multi-Maxwellian distributions of electrons, show that
the strength of the lines is determined by the suprathermal part of
the electron distribution, and the frequency mainly by the thermal
part. For simultaneous up- and downshifted plasma line differences in
intensity we can deduce currents carried by suprathermals, and for
differences in frequency, currents carried by thermals.

The simultaneous up- and downshifted frequencies of the 3 MHz line in
the diffuse aurora show, on average, an upward field-aligned
suprathermal current during the two main periods when they were
detected, 1940-2030~UT and 2250-2300~UT.  In the arcs in general,
there is an indication of downward thermal current as seen from the
altitudes of the 6.5~MHz echoes. Of course, no rule is without
exceptions, and there are cases where one line is much stronger than
the other or the other line is not at all enhanced, indicating strong
currents.

In the full 7-parameter fit of the incoherent scatter spectrum with the
ion line and the both enhanced plasma lines, we obtained a thermal
current with a
suprathermal distribution of electrons consistent with distributions
derived from precipitating fluxes.

\begin{acknowledgements}
  One of the authors (I.H.) was working under a contract from NIPR and
  is grateful to the Director-General of NIPR for the support. We are
  indebted to the Director and staff of \eiscat{} for operating the
  facility and supplying the data. \eiscat{} is an International
  Association supported by Finland (SA), France (CNRS), the Federal
  Republic of Germany (MPG), Japan (NIPR), Norway (NFR), Sweden (NFR)
  and the United Kingdom (PPARC).
\end{acknowledgements}

\bibliographystyle{egs}

\bibliography{strings,a_z,new}

\end{document}